# EMP Needs of Medical Undergraduates in a Saudi Context

Dr. Choudhry Zahid Javid

PhD Applied Linguistics (English for Specific Purposes)

Assistant Professor, Department of Foreign Languages

Taif University, P. O. Box: 888

Taif, KSA.

Mob: 00966-502312949

Email: chzahidj@hotmail.com

ABSTRACT

Needs analysis (NA) is an information gathering process (Nunan, 1988) that is executed to identify the learners' reasons for learning a language so that they might be taught the relevant teaching materials. This study was undertaken at College of Medicine and Medical Sciences (CMMS), Taif University during the academic year 2008 to identify English for Medical Purposes (EMP) needs of the medical undergraduates. Multiple methods of data collection such as exploratory interviews, observations and a questionnaire were used to find out their linguistic needs and the data were triangulated to validate the findings. The findings offered significant insights in selecting suitable course contents to teach English to freshmen medical undergraduates of CMMS. It has been reported that 1) the students lacked the required level of English language proficiency, 2) they needed reading and speaking skills more than others and 3) the data analysis also helped identify the most important language tasks in the context of CMMS. Both populations; the faculty members and the medical undergraduates, recommended that English contact hours should be increased and that EMP should be emphasized more instead of EAP to prepare the learners for their medical studies.

Literature Review

Much research has reported that a comprehensive Needs Analysis (NA) is necessary for all ambitious institutions. Considering its immense importance, a comprehensive NA study was conducted at College of Medicine and Medical Sciences (CMMS) during the second semester of the academic year 2008 to identify English for Medical Purposes (EMP) needs of the medical undergraduates so that suitable and appropriate course contents might be developed to meet the specific needs of these students.

NA is an integral component of all learner-centered, tailor-made and focused course contents (Richards, 1983). Research has provided sufficient insights of its crucial role in systematic curriculum development in the realm of English for Specific Purposes (ESP) and English for Academic Purposes (EAP). The last two decades of the twentieth century have witnessed a revolution in this practice of academic excellence (Hutchinson & Waters, 1987; Coleman, 1988; Richards et al., 1992; Johns & Dudley-Evans, 1991; West, 1997; Edwards, 2000; Richards, 1990; Jordan, 1997; Burvik, 1989; Braine, 2001; Chan, 2001; Leki & Carson, 1994; West, 1994; Detaramani & Chan, 1999). Richards et al. (1992) defined NA as "the process of determining the needs for which a learner or group of learners requires a language" (p. 242). Nunan (1988) has ascribed to it as information-gathering process and stated that NA refers to the "techniques and procedures for collecting information to be used in syllabus design" (p. 13). NA has also been referred to as the process of establishing "what and how" of a course (Dudley-Evans & St. John, 1998).

Nunan (1988) reported that "during the 1970s' NA procedures made their appearance in language planning" and became "widespread in language teaching" (p. 43). The traditional trend of following intuited needs by the language teachers was replaced by a systematic NA and it was rightly stated that "whereas English had previously decided its own destiny; it now became subject to the wishes, needs and demands of people other than language teachers" (Hutchinson & Waters, 1987: p. 7).

Identification of learners' specific reasons of learning a language enables English language teaching (ELT) professionals to cater for the specific needs which saves a lot of time and efforts that used to be wasted by teaching irrelevant ELT material to the learners. The last two decades of the 20$^{th}$ century witnessed a revolution in the field of NA. This modern trend has assigned ELT professional the challenging task to meet learners' needs "as learners' requirements determine the content and aims of the teaching, it is vital, therefore, to discover what these requirements are" (Hadley, 2006: p. 3).

The historical work of Munby (1987) initiated a new era of NA in the field of ESP. Braine (2001) reported that before Munby's model, ESP courses were designed according to "teachers' intuitions" of students' academic needs whereas post-Munby ESP literature was full of NA studies that were carried out to identify learners' specific needs that provided the foundation for tailor-made ESP courses. Research has offered deep insights into identifying learning needs from the perspective of various actors of learning process: the learner, the teacher and the institution (Coleman, 1998). West (1987) has proposed a comprehensive NA delineation. According to him, 'target situation analysis' was meant to identify the demands of the target situation. 'Deficiency analysis' was defined as the gap between the learners' initial proficiency and what they needed to achieve at the end of an ESP course.

## Material and Methods

Research has suggested that multiple sources and procedures should be used in NA studies to get authentic results (Hutchinson & Waters, 1987; Brown, 1995; Al-Khatib, 2005; Bosher & Smalkoski, 2002; Long, 2005; Tudor, 1996; West, 1994). Talking about the use of multiple sources to collect authentic data, Brown (1995) stated that "……. multiple sources of information should be used in a needs analysis………although the specific combination appropriate for a given situation must be decided on the site by the needs analysts themselves" (p.52). Long (2005) has advocated the use of different kinds of data as he has pointed out that "it is difficult to overemphasize the likelihood that use of multiple measures, as well as multiple sources, will increase the quality of information gathered" (p. 32). He has declared insiders as better informants than outsiders and suggested triangulation of data generated by these sources and methods. Triangulation, a commonly used procedure in the field of humanities, involves (with many variants),

> "the systematic comparison of interim findings from two or more sources, methods or combinations thereof, and an attempt to validate the researchers' (in this case the needs analysts) interim findings by presenting them to the informants, and\or by seeking confirmation or disconfirmation of the current analysis in the data arrived at from the methods and sources" (Jasso-Aguilar, 1999: p. 3).

### Subjects (Population and Sample)

The total enrolled students of CMMS comprised the sample for this study. The researcher started empirical work for this study from the beginning of the second semester of academic year 2008 by observing content-subject classes taught by the teachers of different nationalities who were all non-native English speakers. Ten faculty members and 34 students (2$^{nd}$ & 3$^{rd}$ year) were selected for structured semi open-ended exploratory interviews (See appendices # 2 & 3). The detail of this sample is given below:

Freshmen = 100 students (n=100)

$2^{nd}$ year Students= 75 students (n=75)

$3^{rd}$ year Students = 75 students (n= 75)

Total = 250 students (n=250)

## Research Questions

In this study, the researcher was primarily interested in the present situation needs (PSN) and target situation needs (TSN) of CMMS medical undergraduates. The purpose of the study was to identify what students need to know to study their English medium content-subject materials effectively and efficiently. The following were the research questions:

1. What is CMMS medical undergraduates' English language proficiency?
2. What are PSN of medical undergraduates of CMMS?
3. What are TSN of medical undergraduates of CMMS?

## Instrumentation and Data Collection

The study used three qualitative tools: exploratory interviews and class observations; and one quantitative tool: a structured questionnaire.

The first tool was an observation of 11 classes by the researcher. These sessions were recorded for future reference. The tool served two purposes: it provided the researcher with an opportunity to find out the EMP needs of the students, and it also lent valuable information that helped the researcher to design the interview protocols and the questionnaire. The second tool was a structured semi-open ended interview protocol for the faculty members. They were asked seven open-ended and two closed ended questions to discover the present situation needs (PSN) and target situation needs (TSN) of the subjects. The third tool was a structured semi-open-ended interview protocol for carefully selected 34 $2^{nd}$ and $3^{rd}$ year students studying at CMMS.

A structured questionnaire was developed based on Basturkmen (1998) that had 27 items (See appendix # 4). The first section of the questionnaire gathered personal information about the subjects. The second section elicited their level of competence in English language proficiency. In the third section, the samples were asked 4-point Likert-scale questions regarding their PSN and TSN. This questionnaire was administered to all the subjects studying at CMMS during the second semester of academic year 2008.

## Data Analysis

                                                                      94

The descriptive statistics of the data collected through different sources were calculated by using Statistical Package for Social Sciences (SPSS 10) whereas the data generated through the classroom observations and open-ended questions were described and explained by the researcher. All the data generated were triangulated to reach the final conclusion.

## Results

The results of one quantitative and three qualitative tools have been discussed separately and then triangulated to validate the findings.

### Classroom Observation

The researcher observed/recorded 11 classes according to a structured classroom observation protocol (See appendix # 1). While selecting the classes for observation, it was ensured that a fair selection was made. This specimen included 6 theoretical lectures, 3 small group discussion sessions, and 2 practical sessions. Each observation lasted for a fixed period of time: 30 minutes. All these recordings were replayed and the following factors were calculated: Student Reading Time (SRT), Student Listening Time (SLT), Student Writing Time (SWT), Teacher Speaking Time (TST), Student Speaking Time (SST) and Arabic Speaking Time (AST) both for the teachers and the students.

Table 1: Total time for different classroom activities

| No | Activity | Time (minutes) | No | Activity | Time (minutes) |
|---|---|---|---|---|---|
| 1 | SRT | 110 | 4 | TST | 88 |
| 2 | SLT | 132 | 5 | SST | *44* |
| 3 | SWT | 13 | 6 | AST | 98 |

Table 1 shows the total time for different factors and the data has offered deep insights into the fact that reading skills is identified an extremely important language skill for their studies at CMMS. SRT remains 110 minutes for their classroom reading whereas their reading time at home or outside the classroom is not included. SLT has been calculated as 132 minutes which is the maximum time for any activity. SST remains half of TST. Another very important finding has been the substantial amount of time, 98 minutes, dedicated to Arabic speaking in the classroom both by the teachers as well as by the students. SWT remains 13 minutes: done only by a few students.

Student Interview

Thirty-four 2$^{nd}$ and 3$^{rd}$ year students, who secured distinction in the content-subject final examinations, were selected to conduct the exploratory interview. Rizk (2006) remarked about the authenticity of interview technique as follows:

> "This technique is believed to be unbiased and clear, and brings about authentic and real data. Unlike the questionnaire which might have unreal answers, unanswered questions, and vague responses that require clarification, the interview technique results in a better understanding of the students' needs" (p. 97).

The subjects were interviewed according to a self-designed structured semi open-ended interview protocol (See appendix # 3). The first two questions were about subjects' background. Three questions; 3, 6 & 8, were closed-ended and their results were computed and analyzed with the help of SPSS 10. Question 4, 5, 7 and 9 were open-ended questions and their results were described and then explained by the researcher.

Third question captured samples' perception about the most important language skill in the context of CMMS and a majority of 58.8% ranked reading skills while 20.6% rated speaking skills as the most important skill for their academic needs. 17.6% selected listening whereas only 2.9% suggested that writing is the most important skill.

Responding to the question that was about the number of printed pages they need to read in a week, the students reported a range of 15 to 200 pages per week. The mean of 65.8 pages per week with a high SD of 43.35 demonstrated vast differences in their reading practices.

The second last item of the interview protocol asked a sensitive question: use of Arabic in the classrooms. All respondents agreed that they use Arabic in classrooms. They further mentioned that they seldom use English when they talk to each other and use of Arabic is also frequent when they talk to their Arab teachers. The reported percentage of Arabic use in the classroom varied from 30% to 90% of the total teaching time. The mean was calculated as 59.2% with a high SD rate (13.9).

Interview item 4 elicited subjects' responses regarding the problematic areas in learning English language. Out of 47 responses, 14 pointed out that they have problems in reading skills, 13 were about speaking skills difficulties and 12 reported problems in listening activities. Only 8 responses reported problems in writing skills and 3 responses identified grammar in this regard.

Responding to question 7, 29 subjects strongly agreed that they have problems in learning medical terminology. The last question elicited their perception regarding improvements in English curriculum. All the samples unanimously declared that more time should be assigned to English course and reading and speaking skills should be emphasized.

### Faculty Interview

Ten content-subject faculty members were randomly selected by the researcher. The subjects were interviewed according to a self-designed structured semi open-ended interview protocol (See appendix # 2). This tool contained 10 questions: $1^{st}$ & $2^{nd}$ were background questions; questions 3, 4, 6 & 8 closed-ended items to be computed and analyzed through SPSS 10; questions 5, 7, 9 & 10 were open-ended questions to be described and explained by the researcher. All these faculty members were non-native English speakers from four different nationalities.

Question 3 elicited the subjects' perception about the most important language skill for their students to accomplish their medical studies at CMMS. 60% reported reading as the most important followed by listening that was reported by 30% faculty members. The rest 10% identified speaking in this regard.

"Are you satisfied with the English proficiency level of your students?" was the next question in the interview protocol and all the faculty members unanimously stated that they were not satisfied with the proficiency level of the medical undergraduates at CMMS.

Table 2: Students' English language proficiency

| Language skills | Mean | Median | Standard Deviation |
|---|---|---|---|
| Listening | 2.7 | 3 | 0.95 |
| Speaking | 2.4 | 2 | 0.84 |
| Reading | 3.2 | 3 | 1.47 |
| Writing | 2.3 | 2 | 0.95 |

Question 6 on the interview protocol required the faculty members to evaluate English language skills of their $2^{nd}$ and $3^{rd}$ year students on a scale of 1-10: 10 represented the maximum while 1 stood for the least on the scale. With regard to evaluating listening skills of their students, 40% raked them at scale 3, 30% selected 2, 20% declared 4, one sample assigned 1 on this scale and the mean was 2.7. Reporting the proficiency level of their students at CMMS in speaking skills, 50% assigned 2, 30% selected 3, 1 and 4 was reported by one subject each. Their responses showed rather more variation when they were asked to rank their students' proficiency in reading skills. A good majority of 40% respondents assigned 3 while 20% selected 2 and the remaining 4 subjects rated their students' proficiency at 1, 4, 5 and 6 respectively. 40% faculty members awarded 2, 30% chose 3, 20% selected 1 and only one respondent ranked the students at 4 in writing skill. The faculty members' rating reported that the proficiency level of the medical undergraduates at CMMS was quite low as detailed in table 2.

Question 5 in the interview protocol elicited the faculty members' perception regarding various problematic areas their students face in English. All these responses were counted. Out of 20 responses, 6 each reported that the students were weak in listening, reading and speaking skills respectively. One response identified writing as a problematic area and 1 response declared grammar in this regard.

Five out of 10 faculty members strongly recommended that medical terminology should be included in English syllabus while the rest 5 recommended it. It was also recommended that pronunciation should be emphasized in particular. All the faculty members unanimously reported that the students need reading skills the most to read relevant reference material from different sources. Speaking and listening were reported other important skills in this regard.

Question 9 sought their response about how important they thought English was for medical undergraduates to carry out their content-subject studies at CMMS. All subjects strongly agreed that it was very important and the following factors have been mentioned in this regard.

    a. "Their medium of instruction is English.
    b. They need to read their medical textbooks which are in English.
    c. They need to read a lot of reference material in English.
    d. They need to interact with their non-Arab teachers.
    e. They need to interact with the medical professionals in field.

   f.  They need to attend certain medical seminars, workshops and conferences.
   g.  They need proficiency in English to perform better in their different examinations".

The last item of the interview sought their suggestions regarding the improvement in the English course at CMMS. All the respondents proposed that the students should be given maximum practice in all language skills and that reading, listening and speaking skills should be emphasized. Ten responses recommended teaching pronunciation of important medical terminology in the English course. Nine proposed that EMP should be taught instead of EAP. Nine responses suggested that more time should be assigned to English course during the first year. Four responses recommended that special English language courses should be arranged for the students by the administration of CMMS in the evening.

Student Questionnaire: Linguistic Needs

In this section, the researcher dealt with the analysis of the data collected through the student questionnaire (See appendices 4&5). Questions 1 to 5 elicited subjects' perception of their level of proficiency in different English language skills. Questions 6 to 27 attempted to thoroughly capture subjects' perceived English language needs for effectively carrying out their studies at CMMS. The researcher visited students in their classes and personally administered the questionnaire to all the subjects. The researcher got 236 questionnaires filled in. The following format was observed to record subjects' responses.

1. 5-point scale (items 1-5):

   1. excellent          2. very good          3. good
   4. fair               5. poor

2. 4-point scale (items 6-27):

   1. very important                    2. important
   3. less important                    4. not important

Table 3: Means and standard deviations for all questionnaire items

| No | Labels | n | Mean | SD |
|---|---|---|---|---|
| 1 | Level of proficiency you have in listening skill. | 236 | 2.58 | 1.09 |
| 2 | Level of proficiency you have in speaking skill. | 236 | 3.11 | 1.08 |
| 3 | Level of proficiency you have in reading skill. | 236 | 2.26 | 1.04 |

| 4 | Level of proficiency you have in writing skill. | 236 | 2.54 | 1.19 |
|---|---|---|---|---|
| 5 | Level of proficiency you have in grammar. | 236 | 2.88 | 1.24 |
| 6 | How important is listening skill for their studies at CMMS? | 228 | 1.76 | 0.81 |
| 7 | How important is speaking skill for their studies at CMMS? | 228 | 2.03 | 0.96 |
| 8 | How important is reading skill for their studies at CMMS? | 228 | 2.48 | 0.91 |
| 9 | How important is writing skill for their studies at CMMS? | 228 | 3.73 | 0.56 |
| 10 | How important is grammar for their studies at CMMS? | 235 | 2.03 | 0.81 |
| 11 | How important is listening to lectures? | 236 | 1.22 | 0.51 |
| 12 | How important is understanding instructions? | 236 | 1.49 | 0.69 |
| 13 | How important is following question/answer sessions? | 233 | 1.74 | 0.74 |
| 14 | How important is understanding power point presentations? | 235 | 1.46 | 0.71 |
| 15 | How important is asking questions? | 235 | 1.37 | 0.54 |
| 16 | How important is participating in discussions? | 235 | 1.39 | 0.60 |
| 17 | How important is answering the questions? | 235 | 1.50 | 0.66 |
| 18 | How important is giving oral presentations? | 235 | 1.83 | 0.81 |
| 19 | How important is interacting with doctors in field? | 235 | 1.40 | 0.62 |
| 20 | How important is reading textbooks? | 235 | 1.25 | 0.47 |
| 21 | How important is reading articles in journals? | 235 | 2.38 | 0.79 |
| 22 | How important is reading handouts given by teachers? | 235 | 1.60 | 0.69 |
| 23 | How important is reading instructions for assignments? | 235 | 1.60 | 0.65 |
| 24 | How important is taking notes during lectures? | 235 | 1.26 | 0.55 |
| 25 | How important is writing for class quizzes and exams? | 235 | 1.81 | 0.82 |
| 26 | How important is writing assignment and homework? | 235 | 1.66 | 0.72 |
| 27 | How important is writing certain reports? | 235 | 1.91 | 0.85 |

First 5 questions of this questionnaire elicited subjects' responses regarding their perceived level of proficiency in different English language skills. Reading was ranked 1st with mean value of 2.26 that was followed by writing. Listening was perceived 3rd, grammar was found 4th and speaking was 5th. Questionnaire items 6-10 sought subjects' responses regarding the importance of various language skills for their studies at CMMS. A high mean value of 1.76 reported that they needed listening the most followed by speaking. Reading was identified 3rd and writing was considered the least important.

Questions 11 to 14 required the samples to respond to different listening tasks at CMMS and "listening to lectures" has been declared the most important listening task. "Understanding power point presentations" was the 2nd on this ranking followed by "understanding instructions" and "following question/answer sessions respectively. The next 5 questions (15-19) were about various speaking tasks and "asking questions" was reported the most important speaking task with mean value of 1.37 that was followed by "participating in discussions" and "interacting with doctors" with no significant difference in their mean values. "Answering questions" and "interacting with doctors" was assessed the least important speaking tasks. As far as reading tasks were concerned, high mean value of 1.25 was assigned to "reading textbooks" and it was followed by "reading journals". Same mean (1.6) was calculated for the remaining two reading tasks: "reading handouts" and "reading instructions". The last 4 items investigated subjects' responses about important writing tasks and they ranked "taking notes during classes" as the most important writing skills task whereas "writing reports" was declared as the least important one. "Writing assignments and homework" was identified 2nd and "writing classroom quizzes and examinations" was reported 3rd on this ranking.

## Discussion

### Subjects' English Language Proficiency

The following were the findings of students' self-reported evaluation of their proficiency in different skills in English Language. The majority of the students, i.e., more than 50% evaluated their proficiency level as excellent or very good. Unlike students' self-evaluation of their higher proficiency level, the faculty members ranked them between 1 to 4 on a scale of 1-10: 1 indicated the minimum level of proficiency while 10 stood for the maximum. Classroom observations as well as the researcher's personal interaction with the students in the capacity of their English teacher for the last three years supported faculty members' opinion that English proficiency level of the majority of the students was low. Triangulation of the results suggested that the students studying at CMMS did not have the required level of English language proficiency and that their reported proficiency was much higher as compared to their actual proficiency. Much research has reported that Saudi students' reported proficiency in different language skills is exaggerated (AlHarbi, 2005; Al-Gorashi, 1988; Al-mulhim, 2001; Basturkmen, 1998; Ghenghesh, Hamed & Abdelfattah, 2011). This low proficiency negatively affects their content-subject studies as well. Eissa, Misbah & Al- Mutawa (1988) stated that English as medium of instruction creates serious problems for science-major Kuwaiti students: the low proficiency adversely affects their comprehension of scientific concepts and motivation for learning. The exaggerated evaluation of their English language proficiency seems to create another psychological and pedagogical issue that Saudi students prefer to be taught in Arabic to avoid English as medium of instruction. This finding is in line with Assuhaimi & Al-Barr's (1992) study reporting that medical students at King Faisal University in Saudi Arabia stated that one third or more time is saved when they read or write in Arabic. It has also been reported that 66% of the students at King Saud University wanted to have Arabic as medium of instruction, 57% wanted Arabic textbooks, 53% wanted to write projects in Arabic and 33% wanted to have exams in Arabic (Al-Muhandes & Bakri, 1998).

## Importance of English Language Skills

The data generated by the 4 tools; the questionnaire, student interview, faculty interview and class observations, identified writing skills as the least important skills confirming Basturkmen's (1998) findings that writing was not very important for freshmen students of Kuwait University but reported some contradictory findings regarding other skills. The questionnaire data indicated that listening was the most important skill followed by speaking skills. Reading was ranked third. But it seemed that students over-generalized the importance of different language skills without considering their specific academic needs at CMMS. It has been reported that "the questionnaire ----- might have unreal answers, --- vague responses that require clarification" (Rizk, 2006: p. 97). Data generated by the three remaining tools offered uniform findings that reading was the most important skill to carry out their studies effectively at CMMS. The findings confirmed Guo (1987) and Rettanapinyowing et al. (1988) who reported reading as the most important language skill for medical students. Chia et al. (1999) conducted a study to identify EMP needs of Taiwanese medical students and reported that students as well as faculty members declared reading as the most important one for medical studies. The students identified speaking skills as the second most important English language skill whereas the faculty perceived listening as the second most important skill. The findings have partially confirmed Guo's (1987) findings but confirmed Javid (2011) who reported that Saudi medical undergraduates perceived speaking as the second most important skill after reading skills.

The following are the nine most important language tasks according to the findings of the student questionnaire. These are recorded here in order of priority and include: listening to lectures; reading textbooks; taking notes during lectures; interacting with doctors in field; participating in discussions; asking questions; understanding power point presentations; understanding instructions and answering the questions.

## Findings and conclusion

The researcher carefully analyzed and triangulated the data generated from the above-mentioned four research tools that report the following findings. There seems an agreement on the following points that:

- majority of the students lacked required English language proficiency.
- more time should be assigned to English language course.
- reading and speaking skills should be emphasized.
- the nine important language tasks reported by the subjects should be emphasized.
- medical terminology should be taught in English language course.
- use of Arabic should be prohibited in the classrooms.
- EMP should be emphasized more instead of EAP.

Javid, C. Z. (2011). EMP Needs of Medical Undergraduates in a Saudi Context. *Kashmir Journal of Language Research*, 14(1), 89-110.

Appendix # 1

Classroom Observation Protocol

1. Subject:
2. Date:
3. Time:
4. Name of the Teacher:
5. Time for different activities:

| o | Activity | Time (minutes) | No | Activity | Time (minutes) |
|---|---|---|---|---|---|
| 1 | SRT | | 4 | TST | |
| 2 | SLT | | 5 | SST | |
| 3 | SWT | | 6 | AST | |

6. Any other observations:

_______________________________________________________________
_______________________________________________________________
_______________________________________________________________
_______________________________________________________________
_______________________________________________________

Appendix # 2

Faculty Members' Interview Protocol

1. Name:
2. What subject(s) do you teach?
3. What language skill do the students need the most to carry out their studies effectively at CMMS?
4. Are you satisfied with their level of English?
5. What are the language areas where your students feel problems?
6. How would you rank your $2^{nd}$ and $3^{rd}$ year students? Rank them on a scale of 1 to 10.

|  | 1 | 2 | 3 | 4 | 5 | 6 | 7 | 8 | 9 | 10 |
|---|---|---|---|---|---|---|---|---|---|---|
| Listening |  |  |  |  |  |  |  |  |  |  |
| Speaking |  |  |  |  |  |  |  |  |  |  |
| Reading |  |  |  |  |  |  |  |  |  |  |
| Writing |  |  |  |  |  |  |  |  |  |  |

7. Do they need to be taught medical terminology in English course? Why?

8. What language skills do they need the most to deal with small group discussion task?

9. How important is English to carry out their studies effectively at CMMS? Please give reasons.

10. What do you think English course must offer to prepare the students to ensure maximum benefit?

  

Appendix # 3

Students' Interview Protocol

1. Name:

2. Year:

3. What language skill do you need the most to carry out your studies at CMMS?

4. What are the language areas where you feel most problems?

5. How many pages of printed material do you read in a week?

6. What problems do you face in learning medical terms?

7. How often do you use Arabic in classroom? Give percentage of its use.

8. What improvements do you suggest in English curriculum?

   109

Appendix # 4

Student Questionnaire (English)

I- PERSONAL
  Name:                                    Mobile:
  Date:                                    Email:

II- BACKGROUND

  * What level of proficiency do you think you have in the following language skills?
              Excellent   very good   good   fair   poor
1. Listening   ____        ____        ____   ____   ____
2. Speaking    ____        ____        ____   ____   ____
3. Reading     ____        ____        ____   ____   ____
4. Writing     ____        ____        ____   ____   ____
5. Grammar     ____        ____        ____   ____   ____

III- LANGUAGE NEEDS AT CMMS

  * Rank the following according to their importance. Circle the most appropriate choice.
  1 = very important           2 = important
  3 = not important            4 = not applicable

  6. How important is listening skill?          1   2   3   4
  7. How important is speaking skill?           1   2   3   4
  8. How important is reading skill?            1   2   3   4
  9. How important is writing skill?            1   2   3   4
  10. How important is grammar?                 1   2   3   4

  Listening
  11. Listening to lectures                     1   2   3   4
  12. Understanding instructions                1   2   3   4
  13. Following question/answer sessions        1   2   3   4
  14. Understanding power point presentations   1   2   3   4

  Speaking
  15. Asking questions                          1   2   3   4
  16. Participating in discussions              1   2   3   4
  17. Answering the questions                   1   2   3   4
  18. Giving oral presentations                 1   2   3   4
  19. Interacting with doctors in field         1   2   3   4

  Reading
  20. Textbooks                                 1   2   3   4
  21. Articles in journals                      1   2   3   4

| | | | | |
|---|---|---|---|---|
| 22. Handouts given by teachers | 1 | 2 | 3 | 4 |
| 23. Instructions for assignments | 1 | 2 | 3 | 4 |

Writing

| | | | | |
|---|---|---|---|---|
| 24. Taking notes in lectures | 1 | 2 | 3 | 4 |
| 25. Class quizzes and exams | 1 | 2 | 3 | 4 |
| 26. Assignments | 1 | 2 | 3 | 4 |
| 27. Certain reports | 1 | 2 | 3 | 4 |